\documentclass[prd,twocolumn,nofootinbib,preprintnumbers,showpacs]{revtex4-1}

\usepackage{amsfonts}
\usepackage{amsmath}
\usepackage{amssymb}
\usepackage{bm}
\usepackage{dcolumn}
\usepackage{graphicx}
\usepackage{graphics}
\usepackage[latin1]{inputenc}
\usepackage{latexsym}
\usepackage{rotating}
\usepackage{url}
\usepackage{xspace} 
\usepackage[usenames]{color}
\usepackage{mathrsfs}
\usepackage{hyperref}
\usepackage{epstopdf}

\definecolor {darkgreen}{rgb}{0.2,0.7,0.2}

\newcommand\be{\begin{equation}}
\newcommand\ba{\begin{eqnarray}}
\newcommand\ee{\end{equation}}
\newcommand\ea{\end{eqnarray}}

\newcommand\bw{\begin{widetext}}
\newcommand\ew{\end{widetext}}

\newcommand{\nn}{\nonumber}

\newcommand{\mat}{{\mbox{\tiny mat}}}
\newcommand{\eff}{{\mbox{\tiny eff}}}

\newcommand{\hor}{{\mbox{\tiny H}}}
\newcommand{\hork}{{\mbox{\tiny {H,K}}}}

\newcommand{\K}{{\mbox{\tiny K}}}
\newcommand{\ergo}{{\mbox{\tiny ergo}}}
\newcommand{\ergok}{{\mbox{\tiny{ergo,K}}}}

\newcommand{\ISCO}{{\mbox{\tiny ISCO}}}

\begin{document}
\title{Slowly rotating black holes in Einstein-Dilaton-Gauss-Bonnet gravity: \\ Quadratic order in spin solutions}

\author{Dimitry Ayzenberg}
\author{Nicol\'as Yunes}
\affiliation{Department of Physics, Montana State University, Bozeman, MT 59717, USA.}

\date{\today}

\begin{abstract} 

We derive a stationary and axisymmetric black hole solution in Einstein-Dilaton-Gauss-Bonnet gravity to quadratic order in the ratio of the spin angular momentum to the black hole mass squared. This solution introduces new corrections to previously found nonspinning and linear-in-spin solutions. The location of the event horizon and the ergosphere are modified, as well as the quadrupole moment. The new solution is of Petrov type I, although lower order in spin solutions are of Petrov type D. There are no closed timelike curves or spacetime regions that violate causality outside of the event horizon in the new solution. We calculate the modifications to the binding energy, Kepler's third law, and properties of the innermost stable circular orbit. These modifications are important for determining how the electromagnetic properties of accretion disks around supermassive black holes are changed from those expected in general relativity.

\end{abstract}

\pacs{04.50.Kd, 04.70.-s, 04.80.Cc, 04.30.-w}
\preprint{NSF-KITP-14-048}

\maketitle

\section{Introduction}
\allowdisplaybreaks[4] 

The mass of Sagittarius A* (Sgr A*), the supermassive black hole in the center of the Milky Way galaxy, is known to about 10\% uncertainty~\cite{Gillessen:2008qv,Ghez:2008ms}. Due to past technological limitations, mass was the only property that could be inferred from the observation of the orbital motion of nearby stars. The next generation of upgrades to telescopes used in very long baseline interferometers will allow for the determination of other important properties, such as the location of the event horizon and the innermost stable circular orbit (ISCO) from observations of the black hole (BH) shadow and accretion disk, respectively~\cite{Kendrew:2012dm, Gillessen:2010ei, Woillez:2012nu, Stone:2012cf, Bartko:2010kd, Eisner:2010pu, Pott:2008jy, 2009ApJ...697...45B, 2011ApJ...735..110B, Broderick:2011vt, Fish:2010wu, Fish:2009va}. One other property that we wish to infer is whether Sgr A* satisfies the so-called \emph{Kerr hypothesis}, i.e.~that the massive compact objects at the center of galaxies are Kerr BHs. The Kerr metric is the external spacetime of a vacuum, stationary, and axisymmetric BH in general relativity (GR)~\cite{1975PhRvL..34..905R, 1967PhRv..164.1776I, 1968CMaPh...8..245I, 1971PhRvL..26.1344H, 1972CMaPh..25..152H, 1971PhRvL..26..331C}. Modified theories of gravity that may or may not satisfy the Kerr hypothesis, and thus, observations of Sgr A* allow us to test them. 

A modified gravity theory that does not satisfy the Kerr hypothesis is Einstein-Dilaton-Gauss-Bonnet (EDGB) gravity. EDGB modifies the Einstein-Hilbert action through a dynamical scalar field coupled to the Gauss-Bonnet invariant. BHs in EDGB are not described by the Schwarzschild or Kerr metric, and thus, this theory violates the Kerr hypothesis. Instead, BHs acquire corrections that modify important properties, such as the location of the event horizon and the ISCO, relative to the Kerr expectation. EDGB is a well-motivated theory, for example arising from a four-dimensional compactification and low-energy expansion of heterotic string theory, wherein the scalar field is the dilaton~\cite{Moura:2006pz,lrr-2013-9}. In this context, EDGB should be viewed as an \textit{effective field theory} valid up to a cutoff energy scale above which higher order operators cannot be neglected. If the theory is not treated as effective, instabilities can be nonlinearly generated~\cite{Motohashi:2011pw}, which would render the theory ill posed.

Numerical solutions for rapidly rotating BHs in EDGB gravity have been found in~\cite{Kleihaus:2011tg,Torii:1996yi,Kanti:1995vq,Alexeev:1996vs}, but did not treat EDGB as an effective field theory. Early analytic BH solutions in theories motivated by string theory were found and studied in~\cite{Campbell:1992hc,Campbell:1991kz,Campbell:1990ai,Campbell:1990fu}. More recently, analytic BH solutions in EDGB gravity were found in~\cite{Yunes:2011we,Pani:2011gy}. Our work focuses on purely analytic solutions.  Reference~\cite{Yunes:2011we} found an exact, stationary and spherically symmetric solution that represents nonspinning BHs. Reference~\cite{Pani:2011gy} found an approximate, stationary and axisymmetric solution that represents a slowly rotating BH to leading order in the ratio of the spin angular momentum to the BH mass squared. In both cases, the EDGB metrics differed from the Kerr one by modifying certain key properties of BH spacetimes, such as the location of the event horizon and ergosphere. Nonetheless, both solutions were found to be of Petrov type D, just as the Kerr metric. 

In this paper, we find an approximate, slowly rotating BH solution in EDGB gravity to quadratic order in the ratio of the spin angular momentum to the BH mass squared. To derive this solution, we use a new BH perturbation theory method~\cite{PhysRevD.2.2141,Sago:2002fe}, first employed in~\cite{Yagi:2012ya} in the context of modified gravity theories. We treat the second-order-in-spin correction to the EDGB metric as a perturbation away from the leading-order-in-spin one of~\cite{Pani:2011gy}. The perturbation then satisfies a system of differential equations that we decouple through a tensor spherical harmonic decomposition. We finally verify the solution by reinserting it into the field equation and using symbolic manipulation software. Both here and in~\cite{Yunes:2011we,Pani:2011gy}, we work in a \emph{small-coupling approximation}, i.e.~we assume the EDGB  modifications to GR are small and controlled by a dimensionless coupling constant. Such an approximation is consistent with the fact that EDGB is an effective theory, derived from a leading order truncation in the couplings of a more fundamental theory. Thus, its action and associated field equations are only valid to leading order in the coupling.

We then use this solution to study properties of the spacetime. 
First, we establish that the new solution truly represents a black hole, i.e.~that it contains a singularity that is hidden inside an event horizon, and we compute the shift in the location of the event horizon and the ergosphere. Such a study was not possible with the linear-in-spin solution of~\cite{Pani:2011gy}, since it requires quadratic-in-spin corrections to the metric. 
Second, we show that no closed timelike curves exist and that the signature of the metric does not flip outside of the event horizon. This helps justify the perturbative construction of the solution, as small GR deformations should not lead to large modifications in the causal structure of spacetime. 
Third, we find that the quadratic-order-in-spin corrections force the new solution to be of Petrov type I. This is in contrast to black hole solutions in GR and the nonspinning and linear-in-spin black hole solutions in~\cite{Yunes:2011we} and~\cite{Pani:2011gy}, all of which are of Petrov type D. Knowledge of the Petrov type may aid in the construction of analytic black hole solutions that are rapidly spinning. 
Finally, we study the behavior of test particles in orbit around the new EDGB black hole, by obtaining corrections to the orbital binding energy, the angular momentum, the orbital frequency, and the ISCO frequency, and we compute the deformation to the quadrupole moment of the spacetime. All of this could aid in constraining EDGB observationally in the future with electromagnetic~\cite{lrr-2008-9} or gravitational wave observations~\cite{lrr-2013-9}.

The remainder of this paper presents the details pertaining to these results. Section~\ref{sec:edgb} gives a brief summary of EDGB gravity. Section~\ref{sec:rot-bh} first describes the approximation scheme used to find BH solutions and then describes the solutions found in~\cite{Yunes:2011we} and~\cite{Pani:2011gy} and the new solution found in this paper. Section~\ref{sec:props} studies the basic properties of the new solution, such as the location of the event horizon and ergosphere. Section~\ref{sec:energy} discusses the properties associated with particles in orbit around the BH, such as the ISCO and curves of zero velocity. Section~\ref{sec:concs} concludes by summarizing the results, discussing the observational implications, and proposing possible future research.

Throughout we use the following conventions: the metric signature $(-,+,+,+)$; latin letters in index lists stand for spacetime indices; parentheses and brackets in index lists for symmetrization and antisymmetrization, respectively, i.e. $A_{(ab)}=(A_{ab}-A_{ba})/2$ and $A_{[ab]}=(A_{ab}-A_{ba})/2$; geometric units with $G=c=1$.

\section{EDGB Gravity}
\label{sec:edgb}

This theory is defined by the action
\begin{align}
S\equiv&\int d^4x\sqrt{-g}\left\{\kappa R+\alpha e^{\vartheta}\left[R^2-4R_{ab}R^{ab}+R_{abcd}R^{abcd}\right]\right.
\nonumber \\
&\left.-\frac{\beta}{2}\left[\nabla_a\vartheta\nabla^a\vartheta+2V(\vartheta)\right]+\mathcal{L}\mat\right\}.
\end{align}
Here, $g$ stands for the determinant of the metric $g_{ab}$. $R$, $R_{ab}$, and $R_{abcd}$ are the Ricci scalar, Ricci tensor, and the Riemann tensor. $\mathcal{L}\mat$ is the external matter Lagrangian. $\vartheta$ is a field and $V(\vartheta)$ is an additional potential. $(\alpha,\beta)$ are coupling constants, and $\kappa=1/(16\pi)$. For convenience, we define a dimensionless parameter
\begin{equation}
\zeta=\frac{\alpha^2}{\kappa\beta M^4},\label{cc}
\end{equation}
where $M$ is the typical mass of the system.

We assume $\vartheta$ is small, otherwise $e^{\vartheta}$ becomes large which effectively rescales the coupling constant $\alpha$ to large values and the theory will no longer be effective. Moreover, a large value of $e^{\vartheta}$ would lead to a large modification to GR, which has been ruled out by weak-field tests. Assuming small $\vartheta$, we Taylor expand $e^{\vartheta}=1+\vartheta+\mathcal{O}(\vartheta^2)$ and note the $\vartheta$-independent terms are irrelevant, ie.~they lead to a theory identical to GR because the Gauss-Bonnet invariant is a topological invariant. The field equations are then
\begin{equation}
G_{ab}+\frac{\alpha}{\kappa}\mathcal{D}_{ab}^{(\vartheta)}=\frac{1}{2\kappa}\left(T_{ab}^{\mat}+T_{ab}^{(\vartheta)}\right),\label{fe}
\end{equation}
where
\begin{equation}
T_{ab}^{(\vartheta)}=\beta\left[\nabla_a\vartheta\nabla_b\vartheta-\frac{1}{2}g_{ab}\left(\nabla_c\vartheta\nabla^c\vartheta-2V(\vartheta)\right)\right]
\end{equation}
is the scalar field stress-energy tensor and
\begin{align}
\mathcal{D}_{ab}^{(\vartheta)}&\equiv -2R\nabla_a\nabla_b\vartheta+2\left(g_{ab}R-2R_{ab}\right)\nabla^c\nabla_c\vartheta
\nonumber \\
&+8R_{c(a}\nabla^c\nabla_{b)}\vartheta-4g_{ab}R^{cd}\nabla_c\nabla_d\vartheta
\nonumber \\
&+4R_{acbd}\nabla^c\nabla^d\vartheta.
\end{align}
Notice that the field equations remain of second-order. Variation of the action with respect to $\vartheta$ yields the scalar field equation
\begin{equation}
\beta\square\vartheta-\beta\frac{dV}{d\vartheta}=-\alpha\left(R^2-4R_{ab}R^{ab}+R_{abcd}R^{abcd}\right).\label{sfe}
\end{equation}

Before proceeding, we must make a choice for the potential $V(\vartheta)$. If we chose a nonzero potential, usually a mass for the scalar field would be generated, rendering the field short ranged. But EDGB has a shift symmetry ($\vartheta\rightarrow\vartheta+$const) and theories with such a symmetry do not allow mass terms, rendering the field long ranged. Henceforth, we choose $V(\vartheta)=0$. 

\section{Rotating Black Hole Solutions}
\label{sec:rot-bh}

We use two approximation schemes as set out in~\cite{Yagi:2012ya} to obtain a slowly rotating BH solution in EDGB gravity at quadratic order in spin. To find the second-order-in-spin solution we use the non-spinning and linear in spin solutions found by~\cite{Yunes:2011we} and~\cite{Pani:2011gy}, respectively.

\subsection{Approximation schemes}

Following~\cite{Yunes:2009hc}, we consider stationary and axisymmetric BH solutions in EDGB gravity with small coupling ($\zeta\ll 1$) and 
slow rotation ($\chi\ll 1$). Throughout $m$ is the mass of the BH, $a\equiv S/m$ where $S$ is the magnitude of the spin angular momentum of 
the BH, so that $\chi\equiv a/m$ is dimensionless. The small-coupling approximation treats EDGB modifications as small perturbations to the GR solution.

In the small-coupling approximation, we can expand the full metric as
\begin{equation}
\label{metric-exp}
g_{ab}=g_{ab}^{(0)}+\alpha'^2g_{ab}^{(2)}+\mathcal{O}(\alpha'^4),
\end{equation}
where $\alpha'$ is a bookkeeping parameter that labels the order of the small-coupling approximation, with $g_{ab}^{(n)}\propto\alpha^n$. In the above equation, $g_{ab}^{(0)}$ is the full Kerr metric, while $g_{ab}^{(2)}$ is a deformation of the GR metric to leading order in $\alpha'$. Notice, therefore, that in the GR limit ($\alpha \to 0$ or $\zeta \to 0$), the full metric reduces exactly to the Kerr metric.

We will here work in Boyer-Lindquist-like coordinates $(t,r,\theta,\phi)$, so that we can work with the Kerr metric in the form
\begin{align}
ds_{\K}^2=&-\left(1-\frac{2mr}{\Sigma}\right)dt^2-\frac{4mar\sin^2\theta}{\Sigma}dtd\phi+\frac{\Sigma}{\Delta}dr^2
\nonumber \\
&+\Sigma 
d\theta^2+\left(r^2+a^2+\frac{2ma^2r\sin^2\theta}{\Sigma}\right)\sin^2\theta 
d\phi^2,
\end{align}
with $\Delta\equiv r^2-2mr+a^2$ and $\Sigma\equiv r^2+a^2\cos^2\theta$.

In the slow-rotation approximation, one can reexpand the $\zeta$-expanded metric of Eq.~\eqref{metric-exp}. The Kerr metric, for example, can be expanded in the familiar form
\begin{align}
g_{ab}^{(0)}=& g_{ab}^{(0,0)}+\chi' g_{ab}^{(1,0)}+\chi'^2 
g_{ab}^{(2,0)}+\mathcal{O}(\chi'^3),
\end{align}
where $\chi'$ is another bookkeeping parameter that labels the order of the slow-rotation approximation. The quantity $g_{ab}^{(0,0)}$ is here the Schwarzschild metric, while $g_{ab}^{(1,0)}$ and $g_{ab}^{(2,0)}$ are $\chi'$ perturbations. In this paper, we will expand the GR deformation $g_{ab}^{(2)}$ in the slow-rotation approximation as follows:
\begin{align}
\alpha'^2 g_{ab}^{(2)}= &\alpha'^2 g_{ab}^{(0,2)}+\chi' \alpha'^2  
g_{ab}^{(1,2)}+\chi'^2 \alpha'^2  g_{ab}^{(2,2)}
\nonumber \\
&+\mathcal{O}(\alpha'^2 \chi'^3),\label{mexp}
\end{align}
where note that $g_{ab}^{(i,j)}\propto\chi^i \alpha^j$. Such an expansion is justified from the previous work in~\cite{Yunes:2011we,Pani:2011gy}. Even though we find the GR deformation $g_{ab}^{(2)}$ in a slow-rotation expansion, the Kerr metric part of the full metric can be kept in full $\chi'$-unexpanded form when working on astrophysical applications.

We will also expand the scalar field as follows
\begin{equation}
\label{vartheta-exp}
\vartheta=\alpha'\left[\vartheta^{(0,1)}+\chi'\vartheta^{(1,1)}+\chi'^2\vartheta^{(2,1)}\right]+\mathcal{O}(\alpha'\chi'^3).
\end{equation}
Note that the leading-order term is proportional to $\alpha$, as must be the case from Eq.~\eqref{sfe}. There is no $\mathcal{O}(\alpha'^2)$ term and we have neglected terms of $\mathcal{O}(\alpha'^3)$ as they do not affect the metric perturbation at $\mathcal{O}(\alpha'^2)$.

\subsection{BH solutions to $\mathcal{O}(\alpha'^2\chi^0)$ and $\mathcal{O}(\alpha'^2\chi')$}

Yunes and Stein found that to $\mathcal{O}(\alpha'^2\chi'^0)$~\cite{Yunes:2011we}
\begin{equation}
\label{vartheta01}
\vartheta^{(0,1)}=\frac{\alpha}{\beta}\frac{2}{mr}\left(1+\frac{m}{r}+\frac{4}{3}\frac{m^2}{r^2}\right)
\end{equation}
and the only nonvanishing terms in $g_{ab}^{(0,2)}$ is
\begin{align}
\label{gtt-Order0}
g_{tt}^{(0,2)}=&-\frac{\zeta}{3}\frac{m^3}{r^3}
\nonumber \\
&\times\left[1+\frac{26m}{r}+\frac{66}{5}\frac{m^2}{r^2}+\frac{96}{5}\frac{m^3}{r^3}-\frac{80m^4}{r^4}\right],
\\
g_{rr}^{(0,2)}=&-\frac{\zeta}{f^2}\frac{m^2}{r^2}
\nonumber \\
&\times\left[1+\frac{m}{r}+\frac{52}{3}\frac{m^2}{r^2}+\frac{2m^3}{r^3}+\frac{16}{5}\frac{m^4}{r^4}-\frac{368}{3}\frac{m^5}{r^5}\right],
\label{grr-Order0}
\end{align}
where $f=1-\frac{2m}{r}$ and $\zeta=\frac{\alpha^2}{\beta\kappa m^4}$, with $m$ the BH mass. This mass is the \emph{physical mass}, ie.~that which an observer at infinity would measure for example by observing the motion of stars in orbit around this BH.

Pani et al. found that at $\mathcal{O}(\alpha'^2\chi')$ the scalar field has no correction~\cite{Pani:2011gy}. The only nonvanishing term in $g_{ab}^{(1,2)}$ is
\begin{align}
g_{t\phi}^{(1,2)}=&\frac{3}{5}\zeta m \chi\frac{m^3\sin^2\theta}{r^3}
\nonumber \\
&\times\left[1+\frac{140}{9}\frac{m}{r}+\frac{10m^2}{r^2}+\frac{16m^3}{r^3}-\frac{400}{9}\frac{m^4}{r^4}\right].
\label{gtph-pani}
\end{align}

\subsection{BH solutions at $\mathcal{O}(\alpha'^2\chi'^2)$}

\subsubsection{Scalar field}

The right-hand side of Eq.~\eqref{sfe} is proportional to $\alpha$ and thus, the Gauss-Bonnet invariant need only be expanded to $\mathcal{O}(\alpha'^0)$. Thus, we can substitute the Kerr solution and expand in powers of $\chi'$, noting the first two terms, $R^2$ and $R_{ab}R^{ab}$, vanish and we are left with the Kretchmann scalar:
\begin{align}
R_{abcd}R^{abcd}=&\frac{48m^2}{r^6}-\frac{1008\chi^2m^4\cos^2\theta}{r^8}
\nonumber \\
&+\mathcal{O}(\chi'^4).
\end{align}
The Gauss-Bonnet invariant is a parity even quantity, and as such, it can only depend on even powers of $\chi'$. Thus, $\vartheta^{(n,1)}=0$ for all odd $n$.

Expanding $\square\vartheta$ to $\mathcal{O}(\alpha'\chi'^2)$ and solving Eq.~\eqref{sfe} we find
\begin{align}
\label{vartheta21}
\vartheta^{(2,1)}=&-\frac{\alpha\chi^2}{2\beta m r}\left[1+\frac{m}{r}+\frac{4}{5}\frac{m^2}{r^2}+\frac{2}{5}\frac{m^3}{r^3}\right.
\nonumber \\
&\left.+\frac{28}{5}\frac{m^2\cos^2\theta}{r^2}\left(1+\frac{3m}{r}+\frac{48}{7}\frac{m^2}{r^2}\right)\right].
\end{align}
Our result matches that found in~\cite{Pani:2011gy}.

\subsubsection{Metric tensor}
\label{subsect:metric-tensor}

We can rewrite the expansion of the metric in Eq.~\eqref{mexp} as $g_{ab}=g_{ab}^{(0,0)}+h_{ab}$ where $h_{ab}$ is a metric perturbation away from the Schwarzschild solution. Let us further expand $h_{ab}$ as
\begin{align}
h_{ab}=&\chi'g_{ab}^{(1,0)}+\chi'^2g_{ab}^{(2,0)}+\alpha'^2g_{ab}^{(0,2)}
\nonumber \\
&+\chi'\alpha'^2g_{ab}^{(1,2)}+\chi'^2\alpha'^2g_{ab}^{(2,2)}.
\end{align}
The Einstein tensor can then be expanded as
\begin{equation}
G_{ab}=G_{ab}^{[0]}+G_{ab}^{[1]}+G_{ab}^{[2]}+G_{ab}^{[3]}+\mathcal{O}(h^4),
\end{equation}
where the superscript in square brackets counts the power of $h_{ab}$ that appears in each expression. The Schwarzschild metric satisfies the vacuum Einstein equations, and so the first term $G_{ab}^{[0]}$ vanishes.

We can split the $\mathcal{O}(\alpha'^2\chi'^2)$ part of the Einstein tensor into two terms
\begin{align}
G_{ab}^{(2,2)}=&G_{ab}^{[1]}\left(g_{ab}^{(2,2)}\right)+G_{ab}^{[2]}\left(g_{ab}^{(1,0)},g_{ab}^{(2,0)},g_{ab}^{(0,2)},g_{ab}^{(1,2)}\right)
\nonumber \\
&+G_{ab}^{[3]}\left(g_{ab}^{(1,0)},g_{ab}^{(0,2)}\right),
\end{align}
where the first term depends on the unknown $g_{ab}^{(2,2)}$ and the second term depends on the known $g_{ab}^{(1,0)}$ and $g_{ab}^{(1,2)}$ only. At $\mathcal{O}(\alpha'^2\chi'^2)$ the field equations can then be rewritten as
\begin{equation}
G_{ab}^{[1]}\left(g_{ab}^{(2,2)}\right)=S_{ab}^{(2,2)},\label{fe22}
\end{equation}
where the source term is simply
\begin{align}
S_{ab}^{(2,2)}\equiv&-G_{ab}^{[2]}\left(g_{ab}^{(1,0)},g_{ab}^{(2,0)},g_{ab}^{(0,2)},g_{ab}^{(1,2)}\right)
\nonumber \\
&-G_{ab}^{[3]}\left(g_{ab}^{(1,0)},g_{ab}^{(0,2)}\right)- 4 \left[\frac{\alpha}{\kappa}R_{acbd}\nabla^c\nabla^d\vartheta\right]^{(2,2)}
\nonumber \\
&+\frac{1}{2\kappa}T_{ab}^{(\vartheta)~(2,2)}\,.
\end{align}

In this form, the field equations resemble the equations of BH perturbation theory \cite{PhysRevD.2.2141,Sago:2002fe}. We can interpret $G_{ab}^{[1]}\left(g_{ab}^{(2,2)}\right)$ as the linear part of the Einstein tensor built from an unknown perturbation $g_{ab}^{(2,2)}$ in a Schwarzschild background $g_{ab}^{(0,0)}$. Since the source term $S_{ab}^{(2,2)}$ can be computed exactly, we can use Schwarzschild BH perturbation theory tools to solve for $g_{ab}^{(2,2)}$.

As outlined in~\cite{PhysRevD.2.2141,Sago:2002fe}, we decompose the metric perturbation $g_{ab}^{(2,2)}$ and the source term $S_{ab}^{(2,2)}$ in tensor spherical harmonics. We need only consider the even-parity sector of the metric perturbation, as terms of $\mathcal{O}(\alpha'^2\chi'^2)$ are obviously parity even. The even-parity sector only contains seven independent metric components. We only consider stationary and axisymmetric solutions, which further reduces the independent components to five as well as allowing us to focus only on the $m=0$ mode in the decomposition. We are left with two gauge degrees of freedom, which we fix by using the Zerilli gauge~\cite{PhysRevLett.24.737}. These conditions leave three independent degrees of freedom, which are used to parametrize the metric perturbation as
\begin{align}
g^{(2,2)}_{ab}=&\sum_{\ell}\left[f(r)H_{0 \ell 0}(r) a^{\ell 0 (0)}_{ab}+\frac{1}{f(r)}H_{2 \ell 0}(r) a^{\ell 0}_{ab}\right.
\nonumber \\
&\left.+\sqrt{2}K_{\ell 0}(r)  g^{\ell 0}_{ab}\right].
\end{align}
and the source term
\begin{align}
S^{(2,2)}_{ab}=&\sum_{\ell}\left[A_{\ell 0}^{(0)}(r) a^{\ell 0 (0)}_{ab} +A_{\ell 0}(r) a^{\ell 0}_{ab}+B_{\ell 0}(r) b^{\ell 0}_{ab}\right.
\nonumber \\
&\left.+G_{\ell 0}^{(2)}(r) g^{\ell 0}_{ab} + F_{\ell 0}(r) f^{\ell 0}_{ab}\right],
\end{align}
where $f(r) = 1 - 2 m/r$ is the Schwarzschild factor and $[a^{\ell 0 (0)}_{ab},a^{\ell 0}_{ab},b^{\ell 0}_{ab},g^{\ell 0}_{ab},g^{\ell 0}_{ab}]$ are tensor spherical harmonics defined in Appendix~\ref{app1:tensor-harmonics}. The radial functions $A_{\ell 0}^{(0)}(r)$, $A_{\ell 0}(r)$, $B_{\ell 0}(r)$, $G_{\ell 0}^{(s)}(r)$, and $F_{\ell 0}(r)$ can be obtained by decomposing the source $S^{(2,2)}_{ab}$ in tensor spherical harmonics, and they are presented explicitly in Appendix~\ref{app1:tensor-harmonics}, being non-vanishing only for $\ell =0$ and $\ell =2$. 

The metric, radial functions $[H_{0 \ell 0},H_{2 \ell 0},K_{\ell 0}]$ are to be determined by solving the expanded modified field equations [Eq.~\eqref{fe22}]. The decomposition turns these equations into a system of coupled ordinary differential equations~\cite{PhysRevD.2.2141,Sago:2002fe}:
\begin{widetext}
\begin{align}
&f^2 \frac{d^2 K_{\ell 0}}{dr^2}+\frac{1}{r}f\left(3-\frac{5m}{r}\right)\frac{dK_{\ell 0}}{dr}-\frac{1}{r}f^2\frac{dH_2^{\ell 0}}{dr}
-\frac{1}{r^2}f\left(H_2^{\ell 0}-K_{\ell 0}\right)-\frac{\ell(\ell+1)}{2r^2}f\left(H_2^{\ell 0}+K_{\ell 0}\right)=-A_{\ell 0}^{(0)},\label{dfq1}
\\
&-\frac{r-m }{r^2f}\frac{dK_{\ell 0}}{dr}+\frac{1}{r}\frac{dH_0^{\ell 0}}{dr}+\frac{1}{r^2f}\left(H_2^{\ell 0}-K_{\ell 0}\right)+\frac{\ell(\ell+1)}{2r^2f}\left(K_{\ell 0}-H_0^{\ell 0}\right)=-A_{\ell 0},\label{dfq2}
\\
&f \frac{d}{dr}\left(H_0^{\ell 0}-K_{\ell 0}\right) +\frac{2m}{r^2}H_0^{\ell 0}+\frac{1}{r}\left(1-\frac{m}{r}\right)\left(H_2^{\ell 0}-H_0^{\ell 0}\right)=\frac{rf}{\sqrt{\ell(\ell+1)/2}}B_{\ell 0},\label{dfq3}
\\
&f \frac{d^2K_{\ell 0}}{dr^2}+\frac{2}{r}\left(1-\frac{m}{r}\right)\frac{dK_{\ell 0}}{dr}-f\frac{d^2H_0^{\ell 0}}{dr^2}-\frac{1}{r}\left(1-\frac{m}{r}\right)\frac{dH_2^{\ell 0}}{dr}
 -\frac{r+m}{r^2}\frac{dH_{0\ell 0}}{dr}+\frac{\ell(\ell+1)}{2r^2}\left(H_0^{\ell 0}-H_2^{\ell 0}\right)=\sqrt{2}G_{\ell 0}^{(s)},\label{dfq4}
\\
&\frac{H_0^{\ell 0}-H_2^{\ell 0}}{2} =\frac{r^2F_{\ell 0}}{\sqrt{\ell(\ell+1)(\ell-1)(\ell+2)/2}}.\label{dfq5}
\end{align}
\end{widetext}
In Eqs.~\eqref{dfq1},~\eqref{dfq2}, and~\eqref{dfq4}, $\ell$ can take the values $0$ or $2$, but in Eqs.~\eqref{dfq3} and~\eqref{dfq5} $\ell$ can only equal $2$.

There is one remaining gauge freedom in the $\ell =0$ mode, which we will use to further simplify Eqs.~\eqref{dfq1}-\eqref{dfq5}. After imposing stationarity and axisymmetry, there are three independent variables associated with the $\ell =0$ mode. One of these leads to a redefinition of the spherical areal radius. We set $K_{00}=0$ to eliminate this variable.

To solve the system of differential equations in Eqs.~\eqref{dfq1}-\eqref{dfq5} we start by solving Eq.~\eqref{dfq1} for $H_{200}$.  Equations~\eqref{dfq2} and~\eqref{dfq4} can then be solved for $H_{000}$. With $H_{000}$ and $H_{200}$, the $\ell =2$ functions can be found, $H_{0\ell 0}$, $H_{2\ell 0}$, and $K_{\ell 0}$. The full solution is presented in Appendix~\ref{app1:tensor-harmonics}. Each function is a sum of a homogeneous and inhomogeneous piece, with the former containing integration constants. We choose these constants by requiring (1) that the metric be asymptotically flat at spatial infinity, and (2) that the mass and (magnitude of the) spin angular momentum associated with the new solution is given by $m$ and $ma$, as measured by an observer at spatial infinity.

The metric at $\mathcal{O}(\alpha'^2\chi'^2)$ is then
\begin{widetext}
\begin{align}
g_{tt}^{(2,2)}=&-\frac{4463}{2625}\zeta\chi^2\frac{m^3}{r^3}\left[\left(1+\frac{m}{r}+\frac{27479}{31241}\frac{m^2}{r^2}-\frac{2275145}{187446}\frac{m^3}{r^3}-\frac{2030855}{93723}\frac{m^4}{r^4}-\frac{99975}{4463}\frac{m^5}{r^5}+\frac{1128850}{13389}\frac{m^6}{r^6}+\frac{194600}{4463}\frac{m^7}{r^7}\right.\right.
\nonumber \\
&\left.\left.-\frac{210000}{4463}\frac{m^8}{r^8}\right)\left(3\cos^2\theta-1\right)-\frac{875}{8926}\left(1+\frac{14m}{r}+\frac{52}{5}\frac{m^2}{r^2}+\frac{1214}{15}\frac{m^3}{r^3}+\frac{68m^4}{r^4}+\frac{724}{5}\frac{m^5}{r^5}-\frac{11264}{15}\frac{m^6}{r^6}+\frac{160}{3}\frac{m^7}{r^7}\right)\right],
\label{gtt}
\\
g_{rr}^{(2,2)}=&-\zeta \frac{\chi^2}{f^3} \frac{m^3}{r^3}\left[\frac{4463}{2625}\left(1-\frac{5338}{4463}\frac{m}{r}-\frac{59503}{31241}\frac{m^2}{r^2}-\frac{7433843}{187446}\frac{m^3}{r^3}+\frac{13462040}{93723}\frac{m^4}{r^4}-\frac{7072405}{31241}\frac{m^5}{r^5}\right.\right.
\nonumber \\
&\left.\left.+\frac{9896300}{13389}\frac{m^6}{r^6}-\frac{28857700}{13389}\frac{m^7}{r^7}+\frac{13188000}{4463}\frac{m^8}{r^8}-\frac{7140000}{4463}\frac{m^9}{r^9}\right)\left(3\cos^2\theta-1\right)\right.
\nonumber \\
&\left.-\frac{r}{2m}\left(1-\frac{m}{r}+\frac{10m^2}{r^2}-\frac{12m^3}{r^3}+\frac{218}{3}\frac{m^4}{r^4}+\frac{128}{3}\frac{m^5}{r^5}-\frac{724}{15}\frac{m^6}{r^6}-\frac{22664}{15}\frac{m^7}{r^7}+\frac{25312}{15}\frac{m^8}{r^8}+\frac{1600}{3}\frac{m^9}{r^9}\right)\right],
\\
g_{\theta\theta}^{(2,2)}=&-\frac{4463}{2625}\zeta\chi^2 \frac{m^{3}}{r^{3}}\left(1+\frac{10370}{4463}\frac{m}{r}+\frac{266911}{62482}\frac{m^2}{r^2}+\frac{63365}{13389}\frac{m^3}{r^3}-\frac{309275}{31241}\frac{m^4}{r^4}-\frac{81350}{4463}\frac{m^5}{r^5}-\frac{443800}{13389}\frac{m^6}{r^6}+\frac{210000}{4463}\frac{m^7}{r^7}\right)
\nonumber \\
&\times r^{2} \left(3\cos^2\theta-1\right),
\\
g_{\phi\phi}^{(2,2)}=&g_{\theta\theta}^{(2,2)}\sin^2\theta.
\label{gphph}
\end{align}
\end{widetext}
where all other metric components are zero. We have checked explicitly that this solution satisfies the field equations (Eq.~\eqref{fe}) to $\mathcal{O}(\alpha'^2\chi'^2)$  using symbolic manipulation software.

\subsubsection{Accuracy of the approximate solution}

The approximate solution we derived in the previous subsections is valid only when $\zeta \ll 1$, where recall that $\zeta$ is proportional to the coupling constants of EDGB theory. For this reason, it should be clear that as $\zeta \to 0$, then EDGB theory reduces to GR, and the approximate black hole solution derived in the paper reduces identically to the Kerr metric. To be precise, when $\zeta \to 0$, then the ${\cal{O}}(\chi^{0})$ GR deformations in Eqs.~\eqref{gtt-Order0} and~\eqref{grr-Order0} vanish, the ${\cal{O}}(\chi)$ deformation in Eq.~\eqref{gtph-pani} vanishes and the new, ${\cal{O}}(\chi^{2})$ deformations in Eqs.~\eqref{gtt}-\eqref{gphph} vanish, reducing the metric in Eq.~\eqref{metric-exp} to the Kerr metric. Recall also that an expansion in $\zeta \ll 1$ is valid because EDGB theory must be treated as an effective theory, as explained in the Introduction. 

The approximate solution here derived is also clearly only valid when $\chi \ll 1$, but how large a value of $\chi$ can the solution tolerate without incurring an error larger than some tolerance $\tau$? The only precise way to find this maximum value would be to compare the ${\cal{O}}(\chi^{2})$-accurate metric to a numerical, exact solution, like those of~\cite{Kleihaus:2011tg,Torii:1996yi,Kanti:1995vq,Alexeev:1996vs}. Lacking those numerical solutions, all we can do is estimate the error from the next terms expected in the $\chi \ll 1$ series. From the structure of the solution, we have here neglected terms of ${\cal{O}}(\chi^{3})$ in the $(t,\phi)$ component of the metric and ${\cal{O}}(\chi^{4})$ in the diagonal components of the metric. More precisely, the terms neglected in the approximate solution should be of the form $\chi^{3} f(r) S(\theta)$ and $\chi^{4} g(r) T(\theta)$. From the study of black holes in dynamical Chern-Simons gravity~\cite{Yagi:2012ya}, we expect $f(r) T(\theta)$ and $g(r) S(\theta)$ to be of order unity on the horizon and at the equator, where they will acquire their largest numerical values. Given this, requiring that the neglected terms be smaller than some threshold $\tau$, one expects the approximate solution to be valid up to roughly 
\be
\chi \lesssim \tau^{1/3}\,, \qquad {\rm{and}}
\qquad
\chi \lesssim \tau^{1/4}
\ee
for the $(t,\phi)$ and diagonal components of the metric, respectively. For concreteness, if one picks $\tau = 10\%$, then $a/M \lesssim 0.46$ and $a/M \lesssim 0.56$ respectively.

We can carry out such an accuracy analysis explicitly in the case of the scalar field. This is because one can systematically solve Eq.~\eqref{sfe} order by order in $\chi$, to find higher-order-in-$\chi$ corrections, which we present in Appendix~\ref{app2:high-o-field}. The error in $\vartheta$ due to not including terms of ${\cal{O}}(\chi^{4})$ and higher is then largest at the event horizon, where it reduces to
\be
\label{error}
\vartheta^{(4,1)} + \vartheta^{(6,1)} + \vartheta^{(8,1)} = - \frac{\alpha}{\beta} \left(\frac{9}{40} \chi^{4} + \frac{91}{384} \chi^{6} + \frac{25}{112} \chi^{8}\right)\,.
\ee
As expected, notice that the leading-order error in $\chi$ is of the form predicted above, i.e.~a term of order unity (9/40 in this case) times $\chi^{4}$. We can evaluate Eq.~\eqref{error} as a function of $\chi$ to find the value of the spin at which the error equals some tolerance $\tau$. Doing so, and setting $\beta = \alpha$ for this estimate, we find
\be
\chi \lesssim \frac{2^{3/4} 5^{1/4}}{3^{1/2}} \tau^{1/4} \left[1 - 
\frac{455}{1296} \frac{5^{1/2}}{2^{1/2}} \tau^{1/2} + {\cal{O}}(\tau)\right]\,,
\ee
where we expanded in the small tolerance parameter $\tau \ll 1$. If we set $\tau = 10\%$, we then find $\chi \lesssim 0.67$, which is consistent with the estimate presented above.

\section{Properties of the Solution}
\label{sec:props}

\subsection{Singularity, horizon, and ergosphere}

The spacetime solution we have found has a true singularity at $r=0$. We determined this by calculating the Kretchmann invariant $R_{abcd}R^{abcd}$:
\allowdisplaybreaks[4]
\begin{widetext}
\begin{align}
R_{abcd}R^{abcd}=&48\frac{m^2}{r^6}\left(1-\frac{21a^2\cos^2\theta}{r^2}\right)-32\zeta\frac{m^3}{r^7}\left(1+\frac{1}{2}\frac{m}{r}+\frac{72m^2}{r^2}+\frac{7m^3}{r^3}+\frac{64}{5}\frac{m^4}{r^4}-\frac{840m^5}{r^5}\right)
\nonumber \\
&-\frac{428448}{875}\zeta\frac{m^4}{r^8}\chi^2\left(1+\frac{104315}{80334}\frac{m}{r}+\frac{593165}{281169}\frac{m^2}{r^2}-\frac{6239885}{160668}\frac{m^3}{r^3}-\frac{3108445}{80334}\frac{m^4}{r^4}-\frac{5959775}{80334}\frac{m^5}{r^5}\right.
\nonumber \\
&\left.+\frac{22532275}{40167}\frac{m^6}{r^6}+\frac{97300}{4463}\frac{m^7}{r^7}-\frac{105000}{4463}\frac{m^8}{r^8}\right)\left(3\cos^2\theta-1\right) 
+16 \zeta\frac{m^3}{r^7}\chi^2 
\left(1+\frac{1}{2}\frac{m}{r}\right.
\nonumber \\
&\left.+\frac{122}{3}\frac{m^2}{r^2}+\frac{19}{3}\frac{m^3}{r^3}+\frac{2453}{3}\frac{m^4}{r^4}+\frac{272}{3}\frac{m^5}{r^5}+\frac{3338}{15}\frac{m^6}{r^6}-\frac{255056}{15}\frac{m^7}{r^7}+\frac{80m^8}{r^8}\right)\,.
\end{align}
\end{widetext}
Note that this quantity clearly diverges \emph{only} at $r=0$ in these coordinates.

This metric also possesses an event horizon, i.e.~a null surface generated by null geodesic generators. Since the normal to the surface $n^{\mu}$ must itself be null, event horizons must satisfy the horizon equation~\cite{Hansen:2013owa}
\be
\label{hor-eq}
g^{\mu \nu} \partial_{\mu} F \partial_{\nu} F = 0\,,
\ee
where $F(x^{\alpha})$ is a level surface function such that $n_{\mu} = \partial_{\mu} F$. Using that the spacetime is stationary, axisymmetric, and reflection symmetric about the poles and the equator, the level surfaces can only depend on radius. Without loss of generality, we then let $F(x^{\alpha}) = r - r_{\hor}$, where $F = 0$ defines the horizon location. This then forces Eq.~\eqref{hor-eq} into $g^{rr} = 0$, which is nothing but $g_{tt}g_{\phi\phi}-g_{t\phi}^2=0$~\cite{Poisson2004}. Solving this equation, we find
\begin{equation}
r_{\hor}=r_{\hork}-\frac{49}{40}\zeta m-\frac{277}{960}\zeta m \chi^2\,,\label{horizon}
\end{equation}
with $r_{\hork} = m + (m^{2} - a^{2})^{1/2}$ the Kerr result. Our results agree to ${\cal{O}}(\chi^{0})$ with those of Yunes and Stein~\cite{Yunes:2011we}. Notice that the ${\cal{O}}(\chi^{2})$ corrections act to further shrink the event horizon relative to its Kerr analogue.

The location of the ergosphere can be found by solving $g_{tt}=0$ for r. We find
\begin{equation}
r_{\ergo}=r_{\ergok}-\frac{49}{40}\zeta m-\frac{277}{960}\zeta m \chi^2\left(1-\frac{850}{277}\sin^2\theta\right),
\end{equation}
with the ergosphere in Kerr given by $r_{\ergok}=m+ (m^2+a^2\cos^2\theta)^{1/2}$. Notice that this time the ${\cal{O}}(\chi^{2})$ term does not have a definite sign, but can either act to shrink or enlarge the ergosphere, depending on the latitude angle $\theta$. 

Note that our choice of homogeneous integration constants in computing the metric depends on how we choose to define the mass $m$ and the reduced spin angular momentum $a$. We choose to define these quantities as measured by an observer at infinity, which leads to the metric presented in Sec.~\ref{subsect:metric-tensor}. The angular velocity and area of the event horizon become modified with these definitions
\begin{align}
\Omega_{\hor}&\equiv-\frac{g_{tt}}{g_{t\phi}}\Big|_{r=r_{\hor}}=\Omega_{\hork}\left(1+\frac{21}{20}\zeta\right),
\\
A_{\hor}&\equiv2\pi\int_{0}^{\pi}\sqrt{g_{\theta\theta}g_{\phi\phi}}|_{r=r_H}d\theta
\nonumber \\
&=A_{\hork}\left[1-\frac{49}{40}\zeta\left(1+\frac{19}{98}\chi^2\right)\right],
\end{align}
where $\Omega_{\hork}=a/\left(r_{\hork}^2+a^2\right)$ and $A_{\hork} = 4 \pi (r_{\hork}^{2} + a^{2})$ are the horizon's angular velocity and area for the Kerr metric.

\subsection{Lorentz signature}

If the Lorentzian signature of the metric is not preserved outside the horizon, our perturbative construction is not well justified. We show here that the signature is preserved for a small coupling constant. We denote the determinant of the new metric as $g$ and the determinant of the Kerr metric as $g\K\equiv -r^2\sin^2\theta\left(r^2+a^2\cos^2\theta\right)+\mathcal{O}(\chi'^3)$. The determinant of the metric is then given by
\begin{widetext}
\begin{align}
\frac{g}{g_{\K}}=&1+\frac{m^2}{r^2}\chi^2\cos^2\theta-\zeta\frac{m^2}{r^2}\left(1+\frac{8}{3}\frac{m}{r}+\frac{14m^2}{r^2}+\frac{128}{5}\frac{m^3}{r^3}+\frac{48m^4}{r^4}\right)+\frac{1}{2}\zeta\frac{m^2}{r^2}\chi^2\left(1+\frac{8284}{875}\frac{m}{r}+\frac{13546}{525}\frac{m^2}{r^2}\right.
\nonumber \\
&\left.+\frac{874372}{18375}\frac{m^3}{r^3}-\frac{1422}{175}\frac{m^4}{r^4}+\frac{26234}{147}\frac{m^5}{r^5}+\frac{16412}{105}\frac{m^6}{r^6}+\frac{5248}{5}\frac{m^7}{r^7}-\frac{1120m^8}{r^8}\right)-\frac{8926}{875}\zeta\frac{m^3}{r^3}\chi^2\left(1+\frac{19865}{8926}\frac{m}{r}\right.
\nonumber \\
&\left.+\frac{323804}{93723}\frac{m^2}{r^2}-\frac{106915}{8926}\frac{m^3}{r^3}-\frac{103475}{31241}\frac{m^4}{r^4}-\frac{205425}{4463}\frac{m^5}{r^5}+\frac{618800}{4463}\frac{m^6}{r^6}-\frac{735000}{4463}\frac{m^7}{r^7}\right)\cos^2\theta.
\end{align}
\end{widetext}

The correction terms fall off rapidly as $r\rightarrow\infty$, so it is important to look at the signature of $g/g_{\K}$ at the horizon $r\hor$:
\begin{align}
\frac{g}{g_{\K}}&=1+\frac{1}{4}\chi^2\cos^2\theta
\nonumber \\
&-\frac{361}{120}\zeta\left[1-\frac{731411}{7075600}\chi^2\left(1-\frac{1420033}{731411}\cos^2\theta\right)\right].
\end{align}
Notice that the term in square brackets is always positive, so the $\zeta$ correction is always negative, which could be a problem for a sufficiently large value of the coupling constant. The correction is at a maximum when $\chi=0$, which means the signature flip does not take place as long as $\zeta\lesssim 0.33$. The strongest current constraints on EDGB come from low-mass x-ray binary observations, $\sqrt{\lvert\alpha\rvert}<1.9\times10^5$cm~\cite{Yagi:2012gp}. Setting $\beta=1$ and using a very low-mass BH with $m = 5 M_{\odot}$, this constraint implies $\zeta \lesssim 0.2$. We then see that current constraints already exclude the region of parameter space in which a Lorentz signature flip could occur. Of course, if the BH mass is small enough, then $\zeta$ will become larger, as it scales with $m^{-4}$, but then the small-coupling approximation would break down. 

\subsection{Closed timelike curves}

Closed timelike curves, if they exist, can be found by solving for the region where $g_{\phi\phi}<0$. The explicit form of $g_{\phi\phi}$ was already presented in Eq.~\eqref{gphph}, where we see that the corrections fall off rapidly as $r^{-3}$ relative to the Kerr value of this metric component. Thus, the corrections are largest at the horizon $r\hor$, where 
\begin{align}
g_{\phi\phi}&=4m^2\sin^2\theta\left\{1-\frac{1}{4}\chi^2\cos^2\theta\right.
\nonumber \\
&\left.-\frac{49}{40}\zeta\left[1-\frac{102673}{180075}\chi^2\left(1-\frac{2041527}{821384}\cos^2\theta\right)\right]\right\}.
\end{align} 
The sign of the correction terms depend on the spin, but for small spin the $\chi^0$ term dominates and the correction is always negative. In this case, we see that $\zeta > 0.8$ for $g_{\phi \phi}$ to vanish. As already argued, such values of $\zeta$ are excluded by current constraint for realistic BH masses, and thus, closed timelike curves do not occur. 

\subsection{Multipolar structure}

Following Thorne~\cite{RevModPhys.52.299}, the multipole moment can be read off by transforming the metric to asymptotically Cartesian and mass-centered (ACMC) coordinates. In these coordinates the multipole moments are defined in a spacetime region asymptotically far from the source. To find the quadrupole moment, the coordinate transformation to ACMC must be done such that $g_{tt}$ and $g_{ij}$ at $\mathcal{O}(r^{-2})$ do not contain any angular dependence. In these coordinates, $g_{tt}$ for a stationary and axisymmetric spacetime can be written as
\begin{equation}
g_{tt}=-1+\frac{2m}{r}+\frac{\sqrt{3}}{2}\frac{1}{r^3}\left[Q_{20}Y^{20}+(\ell=0\text{ pole})\right]+\mathcal{O}(\frac{1}{r^4}).\label{qm}
\end{equation}
$Y^{20}$ is the $(\ell,m)=(2,0)$ spherical harmonic and $Q_{20}$ is the $(m=0)$ quadrupole moment.

The correction in the new metric is at $\mathcal{O}(\alpha'^2\chi'^2)$, so it is not affected by the coordinate transformation. The quadrupole moment in the new solution is then
\begin{equation}
Q_{20}=Q_{20,\K}\left(1+\frac{4463}{2625}\zeta\right)\,,
\end{equation}
where $Q_{20,\K}$ is the Kerr quadrupole moment.

\subsection{Petrov type}

Generic spacetimes can be classified into Petrov types by finding the number of distinct principal null directions (PNDs) $k^a$ of the Weyl tensor $C_{abcd}$~\cite{Stephani2003,Campanelli:2008dv}, where $k^a$ must satisfy
\begin{equation}
k^bk^ck_{[e}C_{a]bc[d}k_{f]}=0.
\end{equation}
This is the same as finding the number of distinct PNDs $l^a$ that make one of the Weyl scalars $\Psi_0=0$, which simplifies to finding the number of distinct roots for b in~\cite{Stephani2003}
\begin{equation}
\Psi_0+4b\Psi_1+6b^2\Psi_2+4b^3\Psi_3+b^4\Psi_4=0.\label{roots}
\end{equation}
The $\Psi$'s are five complex Weyl scalars in an arbitrary tetrad with the restriction that $\Psi_4\neq0$.

The spacetime is said to be \textit{algebraically special} if Eq.~\eqref{roots} has at least one degenerate root, and the following relation holds:
\begin{equation}
I^3=27J^2.\label{PI}
\end{equation}
The quadratic and cubic Weyl quantities $I$ and $J$ are defined by~\cite{Stephani2003}
\begin{align}
I&\equiv\frac{1}{2}\tilde C_{abcd}\tilde C^{abcd}
\nonumber \\
&=3\Psi_2^2-4\Psi_1\Psi_3+\Psi_4\Psi_0,\label{I}
\\
J&\equiv-\frac{1}{6}\tilde C_{abcd}\tilde C^{cd}_{~~ef}\tilde C^{efab}
\nonumber \\
&=-\Psi_2^3+2\Psi_1\Psi_3\Psi_2+\Psi_0\Psi_4\Psi_2-\Psi_4\Psi_1^2-\Psi_0\Psi_3^2,\label{J}
\end{align}
where
\begin{equation}
\tilde C_{abcd}\equiv\frac{1}{4}\left(C_{abcd}+\frac{i}{2}\epsilon_{abef}C^{ef}_{~~cd}\right).
\end{equation}
The spacetime is of Petrov type I if Eq.~\eqref{PI} does not hold. The Kerr BH is known to be of Petrov type D. For a spacetime to be type D Eq.~\eqref{PI} must hold along with the following conditions:
\begin{align}
K=&0,\label{K}
\\
N-9L^2=&0,\label{NL}
\end{align}
where $K$, $L$, and $N$ are
\begin{align}
K&\equiv\Psi_1\Psi_4^2-3\Psi_4\Psi_3\Psi_2+2\Psi_3^3,
\\
L&\equiv\Psi_2\Psi_4-\Psi_3^2,
\\
N&\equiv\Psi_4^2I-3L^2
\nonumber \\
&=\Psi_4^3\Psi_0-4\Psi_4^2\Psi_1\Psi_3+6\Psi_4\Psi_2\Psi_3^2-3\Psi_3^4.
\end{align}

One can find a null tetrad for the no-rotating BH solution in EDGB such that $\Psi_2$ is the only nonvanishing Newman-Penrose scalar. Equations~\eqref{PI},~\eqref{K}, and~\eqref{NL} are then trivially satisfied. Thus, the nonspinning solution found in~\cite{Yunes:2011we} is of Petrov Type D.

For the slowly rotating BH solution in EDGB gravity to linear order in spin~\cite{Pani:2011gy}, we first find a principal null tetrad that is a deformation away from the Kerr principal null tetrad. We then find that Eqs.~\eqref{I},~\eqref{K}, and~\eqref{NL} are all satisfied to $\mathcal{O}(\alpha'^4\chi'^2)$. Thus, we find the slowly rotating solution to $\mathcal{O}(\alpha'^2\chi')$ is also of Petrov Type D. \footnote{For a discussion of the order in the perturbation used to compute the Petrov Type see~\cite{Yagi:2012ya}.} 

For the new BH solution at $\mathcal{O}(\alpha'^2\chi'^2)$ the story is different. We first find a principal null tetrad by adding $\mathcal{O}(\alpha'^2\chi'^2)$ deformations to the null tetrad found in the $\mathcal{O}(\alpha'^2\chi')$ case. We then find that Eq.~\eqref{I} is \emph{not} satisfied to $\mathcal{O}(\alpha'^4\chi'^4)$. Thus, the new metric found in this paper is of Petrov Type I, and breaks symmetries that the $\mathcal{O}(\alpha'^2\chi')$ metric had. This suggests that the exact BH solution should be of Petrov type I.

\section{Properties of Test-Particle Orbits}
\label{sec:energy}

\subsection{Conserved quantities}

The metric found here is stationary and axisymmetric, and thus, it possess a timelike and an azimuthal Killing vector, which imply the existence of two conserved quantities: the energy and the ($z$ component of the) angular momentum.  The definitions of $E$ and $L_z$ lead to
\begin{align}
\dot t=&\frac{Eg_{\phi\phi}+L_z g_{t\phi}}{g_{t\phi}^2-g_{tt}g_{\phi\phi}},
\\
\dot\phi=&-\frac{Eg_{t\phi}+L_z g_{tt}}{g_{t\phi}^2-g_{tt}g_{\phi\phi}},
\end{align}
where the overhead dot represents a derivative with respect to the affine parameter. Substituting the above equations into $u^au_a=-1$, where $u^a$ is the particle's four-velocity, we find
\begin{equation}
g_{rr}\dot r^2+g_{\theta\theta}\dot\theta^2=V_{\eff}(r,\theta;E,L_z),
\end{equation}
where the effective potential is
\begin{equation}
V_{\eff}\equiv\frac{E^2g_{\phi\phi}+2EL_zg_{t\phi}+L_z^2g_{tt}}{g_{t\phi}^2-g_{tt}g_{\phi\phi}}-1.\label{Veff}
\end{equation}

For simplicity, we restrict our attention to equatorial, circular orbits. $E$ and $L_z$ can then be obtained from $V\eff=0$ and $\partial V_{\eff}/\partial r=0$ in the form
\begin{align}
E=&E_{\K}+\delta E,\label{Eexp}
\\
L_z=&L_{z,\K}+\delta L_z.\label{Lzexp}
\end{align}
$E_{\K}$ and $L_{z,\K}$ are the energy and $z$ component of the orbital angular momentum for the Kerr spacetime given by \cite{1972ApJ...178..347B}
\begin{align}
E_{\K}\equiv&\frac{r^{3/2}-2mr^{1/2}+am^{1/2}}{r^{3/4}\left(r^{3/2}-3mr^{1/2}+2am^{1/2}\right)^{1/2}},
\\
L_{z,\K}\equiv&\frac{m^{1/2}\left(r^2-2am^{1/2}r^{1/2}+a^2\right)}{r^{3/4}\left(r^{3/2}-3mr^{1/2}+2am^{1/2}\right)^{1/2}},
\end{align}
where $\phi$ is defined to be positive in the direction of prograde orbits. This implies negative $a$ corresponds to retrograde orbits. The corrections from EDGB are
\begin{widetext}
\begin{align}
\delta E\equiv&-\frac{1}{12}\zeta\frac{m^3}{r^{3/2}\left(r-3m\right)^{3/2}}\left(1+\frac{54m}{r}+\frac{198}{5}\frac{m^2}{r^2}+\frac{252}{5}\frac{m^3}{r^3}-\frac{2384}{5}\frac{m^4}{r^4}+\frac{480m^5}{r^5}\right)
\nonumber \\
&+\frac{23}{20}\zeta\chi\frac{m^{9/2}}{r^2\left(r-3m\right)^{5/2}}\left(1+\frac{492}{23}\frac{m}{r}-\frac{458}{23}\frac{m^2}{r^2}-\frac{8}{23}\frac{m^3}{r^3}-\frac{4272}{23}\frac{m^4}{r^4}+\frac{5760}{23}\frac{m^5}{r^5}\right)
\nonumber \\
&+\frac{205821}{441000}\zeta\chi^2\frac{r^{1/2}m^3}{\left(r-3m\right)^{7/2}}\left(1+\frac{29926}{9801}\frac{m}{r}-\frac{2584229}{68607}\frac{m^2}{r^2}-\frac{317782}{68607}\frac{m^3}{r^3}+\frac{14792212}{205821}\frac{m^4}{r^4}+\frac{207551}{6237}\frac{m^5}{r^5}\right.
\nonumber \\
&\left.+\frac{5757700}{9801}\frac{m^6}{r^6}-\frac{257772890}{205821}\frac{m^7}{r^7}+\frac{4064600}{9801}\frac{m^8}{r^8}-\frac{6499000}{3267}\frac{m^9}{r^9}+\frac{4715200}{1089}\frac{m^{10}}{r^{10}}-\frac{280000}{121}\frac{m^{11}}{r^{11}}\right),
\\
\delta L_z\equiv&-\frac{1}{4}\zeta\frac{m^{5/2}}{\left(r-3m\right)^{3/2}}\left(1+\frac{100}{3}\frac{m}{r}-\frac{30m^2}{r^2}+\frac{16}{5}\frac{m^3}{r^3}-\frac{752}{3}\frac{m^4}{r^4}+\frac{320m^5}{r^5}\right)
\nonumber \\
&+\frac{30}{20}\zeta\chi\frac{m^4}{r^{1/2}\left(r-3m\right)^{5/2}}\left(1+\frac{31}{2}\frac{m}{r}-\frac{47}{3}\frac{m^2}{r^2}+\frac{m^3}{r^3}-\frac{126m^4}{r^4}+\frac{1976}{15}\frac{m^5}{r^5}+\frac{80m^6}{r^6}\right)
\nonumber \\
&+\frac{617463}{441000}\zeta\chi^2\frac{r^2 m^{5/2}}{\left(r-3m\right)^{7/2}}\left(1-\frac{86288}{29403}\frac{m}{r}-\frac{2144627}{205821}\frac{m^2}{r^2}+\frac{924068}{68607}\frac{m^3}{r^3}+\frac{27006916}{617463}\frac{m^4}{r^4}-\frac{18616907}{205821}\frac{m^5}{r^5}\right.
\nonumber \\
&\left.+\frac{49732516}{205821}\frac{m^6}{r^6}-\frac{427757690}{617463}\frac{m^7}{r^7}+\frac{197940200}{205821}\frac{m^8}{r^8}-\frac{12547000}{9801}\frac{m^9}{r^9}+\frac{4715200}{3267}\frac{m^{10}}{r^{10}}-\frac{280000}{363}\frac{m^{11}}{r^{11}}\right).
\end{align}
\end{widetext}
Expanding $E$ and $L_z$ in powers of $m/r$, the leading-order corrections to the binding energy $E_b\equiv E-1$ and $L_z$ are
\begin{align}
E_b=&E_{b,\K}\left[1+\frac{1}{6}\zeta\frac{m^2}{r^2}\left(1-\frac{9801}{1750}\chi^2\right)\right],\label{Eb}
\\
L_z=&L_{z,\K}\left[1-\frac{1}{4}\zeta\frac{m^2}{r^2}\left(1-\frac{9801}{1750}\chi^2\right)\right].
\end{align}
Note that the corrections are of 2PN order [proportional to $(m/r)^2$] relative to the leading-order Kerr terms for the energy and angular momentum respectively. These results agree with those of~\cite{Yunes:2011we} to leading order in $\chi$. 

\subsection{Kepler's third law}

The correction to Kepler's third law for a circular orbit can be found by calculating the orbital angular frequency of a test-particle $\omega\equiv L_z/r^2$,
\begin{equation}
\omega^2=\omega^2_{\K}\left[1-\frac{1}{2}\zeta\frac{m^2}{r^2}\left(1-\frac{9801}{1750}\chi^2\right)\right],\label{omega}
\end{equation}
where $\omega_{\K}^2\equiv m\left(r^{3/2}+am^{1/2}\right)^{-2}$~\cite{1972ApJ...178..347B}.

The expressions above for $E$, $L_z$, and $\omega$ are not gauge invariant. We can obtain gauge invariant relations between $E$ and $\omega$ by expanding Eqs.~\eqref{Eb} and~\eqref{omega} to 2PN order and eliminating $m/r$. The result is
\begin{align}
\omega(E)=&\frac{2\sqrt{2}}{m}|E_b|^{3/2}\left\{1+\frac{9}{4}|E_b|+8\sqrt{2}\chi|E_b|^{3/2}\right.
\nonumber \\
&\left.+\frac{891}{32}\left[1+\frac{64}{297}\chi^2-\frac{32}{891}\zeta\left(1-\frac{9801}{1750}\chi^2\right)\right]|E_b|^2\right\}
\nonumber \\
&+\mathcal{O}\left[|E_b|^{4}\right],
\end{align}
and its inverse
\begin{align}
E(\omega)=&1-\frac{1}{2}(m\omega)^{2/3}+\frac{3}{8}(m\omega)^{4/3}-\frac{4}{3}\chi(m\omega)^{5/3}
\nonumber \\
&+\frac{27}{16}\left[1+\frac{8}{27}\chi^2-\frac{4}{81}\zeta\left(1-\frac{9801}{1750}\chi^2\right)\right](m\omega)^2
\nonumber \\
&+\mathcal{O}\left[(m\omega)^{7/3}\right].
\end{align}
This agrees with the standard PN $E$-$\omega$ relation to $\mathcal{O}(\alpha'^0\chi'^0)$~\cite{lrr-2006-5}.

\subsection{ISCO}

Let us now derive the location of the ISCO in this new spacetime. We do so by substituting Eqs.~\eqref{Eexp} and~\eqref{Lzexp} into Eq.~\eqref{Veff}, and then solving $\partial^2V\eff/\partial r^2=0$ for r. The result is
\begin{align}
r_{\ISCO}&=r_{\ISCO,\K}
-\frac{16297}{9720}\zeta m\left(1+\frac{205982\sqrt{6}}{440019}\chi
\right. 
\nonumber \\
&\left.
-\frac{1167369773}{9702418950}\chi^2\right),
\end{align}
where the Kerr ISCO radius is given by~\cite{1972ApJ...178..347B}
\begin{equation}
r_{\ISCO,\K}\equiv m\left\{3+Z_2-\left[\left(3-Z_1\right)\left(3+Z_1+2Z_2\right)\right]^{1/2}\right\}
\end{equation}
with
\begin{align}
Z_1\equiv&1+\left(1-\chi^2\right)^{1/3}\left[\left(1+\chi\right)^{1/3}+\left(1-\chi\right)^{1/3}\right],
\\
Z_2\equiv&\left(3\chi^2+Z_1^2\right)^{1/2}.
\end{align}
The EDGB correction at $\mathcal{O}(\chi'^0)$ agrees with that found in~\cite{Yunes:2011we}. Note that the radial location of the ISCO is not gauge invariant. For a gauge invariant quantity, we compute the angular orbital frequency at ISCO, $\omega_{\ISCO}$:
\begin{align}
\omega_{\ISCO}&=\omega_{\ISCO,\K}
\nonumber \\
&-\frac{13571\sqrt{3}}{3149280}\zeta\frac{1}{m}\left(1+\frac{129655\sqrt{6}}{122139}\chi+\frac{2740701487}{897721650}\chi^2\right),
\end{align}
where $\omega_{\ISCO,\K}=m^{1/2}\left(r^{3/2}_{\ISCO,\K}+\chi m^{3/2}\right)^{-1}$.

\subsection{Curves of zero velocity}

\begin{figure*}[htb]
\begin{center}
\begin{tabular}{c}
\includegraphics[width=9cm,clip=true]{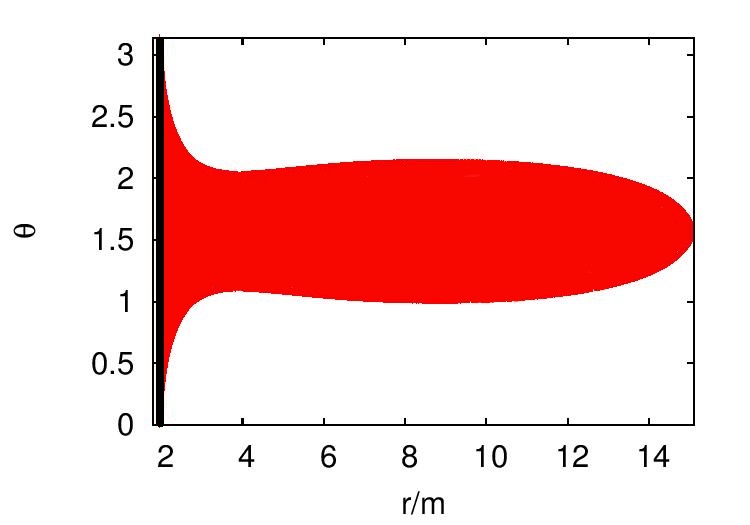}
\includegraphics[width=9cm,clip=true]{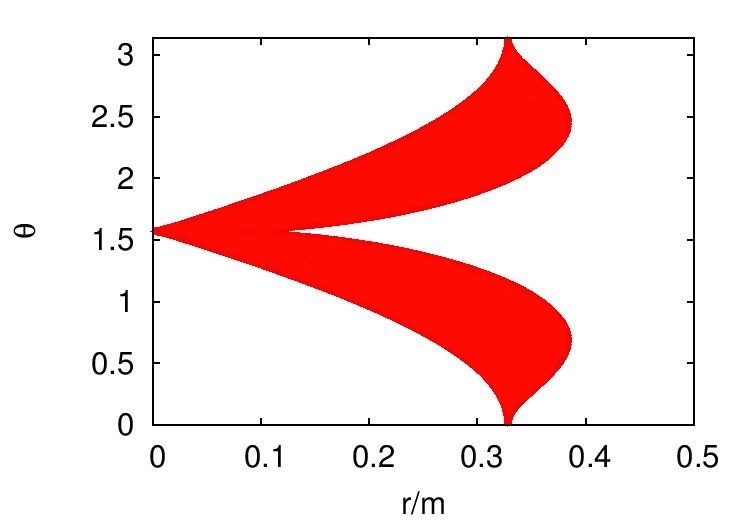}
\\
\includegraphics[width=9cm,clip=true]{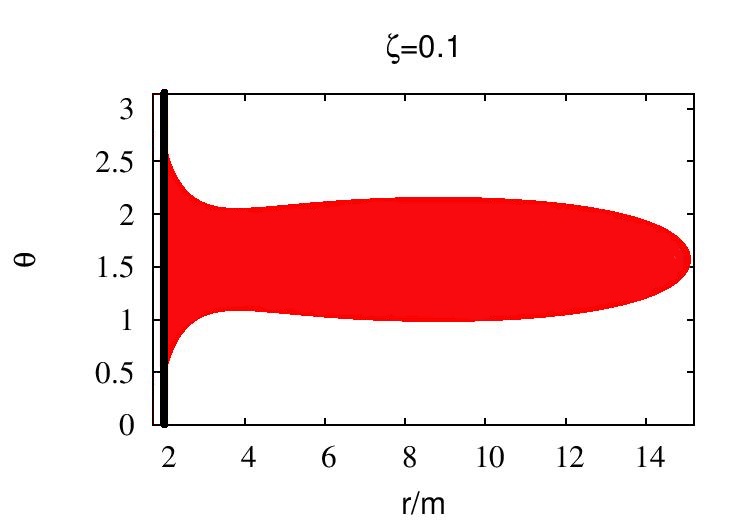}
\includegraphics[width=9cm,clip=true]{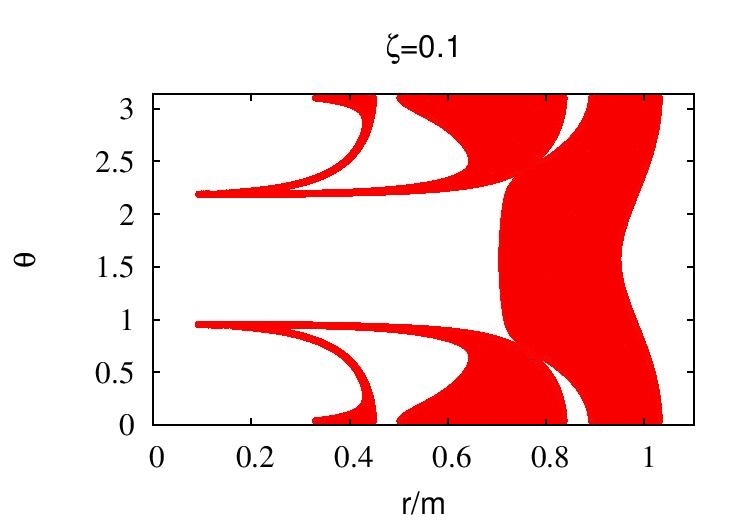}
\end{tabular}
\caption{\label{fig:CZV} Curves of zero velocity ($V_{\eff}=0$) for the Kerr (top) and EDGB (bottom) metric with $\zeta = 0.1$. The red shaded regions show the allowed bound-orbit regions ($V_{\eff}\geq0$) with $E=0.95$, $L_z=3m$, and $\chi=0.3$. The left panels corresponds to the region outside the horizon, while the right ones show the region inside the horizon. The thick black lines at $r/m=1.955$ (top) and  $r/m=1.832$ (bottom) correspond to the location of the horizon for this example.}
\end{center}
\end{figure*}

Last, we will consider curves of zero velocity (CZVs)~\cite{Gair:2007kr,Bambi:2011vc} in the $r$-$\theta$ plane. These curves are where $V_{\eff}=0$ and since the left-hand side of Eq.~\eqref{Veff} is always positive, bound orbits are allowed only if $V_{\eff}\geq0$. Figure~\ref{fig:CZV} shows the CZVs for the Kerr and the new solution. Red shaded regions are where $V_{\eff}\geq0$ and the thick black lines correspond to the location of the event horizon for the particular case considered in the figures. To draw these figures, we expand the metric $g_{ab}$ in the spin parameter $a$ and then calculate $V_{\eff}$.

For both the GR and EDGB case there is one allowed bound-orbit region clearly visible in the region outside of the event horizon. For the region inside the horizon, there is one allowed orbit region in GR, but there are five in the EDGB case. While the regions outside the horizon look similar in GR and EDGB, there are differences not easily visible due to the scale of the figures.The orbits in this outer region are, in principle, distinguishable with gravitational wave observations, as shown in~\cite{Sopuerta:2011te} and~\cite{Canizares:2012is}. The inner regions are drastically different, which is expected as the field is strongest within the horizon and the EDGB corrections modify the strong field regime. However, since these inner regions are within the horizon they cannot be probed with any observations.

\section{Conclusion}
\label{sec:concs}

We found a stationary, axisymmetric BH solution in EDGB gravity in the small-coupling and slow-rotation approximations at linear order in the coupling constant and quadratic order in the spin. The technique used, based on BH perturbation theory, involved decomposing the metric perturbation and source terms in tensor spherical harmonics, which reduced the field equations to a set of coupled, ordinary differential equations. We found new corrections to the metric at quadratic order in spin. We then studied a plethora of properties of this metric, proving that (i) it possesses a curvature singularity inside an event horizon, (ii) the location of the event horizon, ergosphere, horizon area and horizon's angular velocity are all modified relative to the Kerr analogue and (iii) that test-particle orbits in this spacetime are different than those in Kerr due to corrections in the orbital binding energy, angular momentum and effective potential.

As the method used is not specialized to quadratic order in spin and linear order in the coupling constant, an obvious extension of this work is to find solutions to higher order in spin and/or higher order in the coupling constant. In the case of EDGB, however, as it is a linear-order truncation in the coupling constant of a more fundamental theory, any solution is only valid to linear order in the coupling constant.

An interesting and nontrivial property of the new solution is that it is of Petrov type I. This is especially interesting because to zeroth and linear order in spin the solution remains of Petrov type D, and because the Kerr metric is of Petrov type D to all orders in spin. This suggests that the full, exact solution must also be of Petrov type I. Petrov type I spacetimes do not possess a second-order Killing tensor or a Carter-like constant. This implies that geodesic motion may by chaotic once corrections of $\mathcal{O}(\alpha'^2\chi'^2)$ are included. Future work could study whether geodesics in this new metric are chaotic, specifically if there exist chaotic orbits outside of the event horizon.

The new metric solution as well as its properties are important in determining the properties of electromagnetic radiation from accretion disks around a BH. Observations of the electromagnetic radiation near observable BHs, such as Sgr A*, can be a powerful way to test GR~\cite{lrr-2008-9}. An avenue of study would be to determine how observables, such as BH shadows~\cite{Falcke:2013ola} and strong lensing~\cite{Chen:2012kn}, are modified if the BH is described by the new solution found in this paper. Of course, the metric derived here would be appropriate for such tests if and only if the black hole observed has a sufficiently small spin, roughly $S^{2}/M^{4} \lesssim 0.5$. For other, more rapidly rotating black holes, either numerical solutions would have to be used or a higher-order-in-spin approximate solution would have to be derived.

\acknowledgements

We thank Kent Yagi for useful discussions. N.Y. acknowledges support from NSF Grant PHY-1114374 and the NSF CAREER Grant No. PHY-1250636, as well as support provided by the National Aeronautics and Space Administration from Grant No. NNX11AI49G, under sub-award 00001944. This research was supported in part by the National Science Foundation under Grant No. NSF PHY11-25915. Some calculations used the computer algebra systems MAPLE, in combination with the GRTensor II package~\cite{GRT}. 

\appendix
\section{Tensor Harmonics}
\label{app1:tensor-harmonics}

In this paper, we used the following tensor spherical harmonics to decompose the metric perturbation and the source term~\cite{PhysRevD.2.2141,Sago:2002fe}
\begin{equation}
a^{\ell 0 (0)}_{ab} = \begin{pmatrix}
Y_{\ell 0} & 0 & 0 & 0 \\
0 & 0 & 0 & 0 \\
0 & 0 & 0 & 0 \\
0 & 0 & 0 & 0 \\
\end{pmatrix}\,, 
\end{equation}

\begin{equation}
a^{\ell 0}_{ab} = \begin{pmatrix}
0 & 0 & 0 & 0 \\
0 & Y_{\ell 0}  & 0 & 0 \\
0 & 0 & 0 & 0 \\
0 & 0 & 0 & 0 \\
\end{pmatrix}\,,
\end{equation}

\begin{equation}
b^{\ell 0}_{ab} = \frac{r}{\sqrt{2\ell (\ell +1)}} \begin{pmatrix}
0 & 0 & 0 & 0 \\
0 & 0  & \frac{\partial}{\partial \theta} Y_{\ell 0} & 0 \\
0 & \frac{\partial}{\partial \theta} Y_{\ell 0} & 0 & 0 \\
0 & 0 & 0 & 0 \\
\end{pmatrix}\,,
\end{equation}

\begin{equation}
g^{\ell 0}_{ab} = \frac{r^2}{\sqrt{2}} \begin{pmatrix}
0 & 0 & 0 & 0 \\
0 & 0  & 0 & 0 \\
0 & 0 & Y_{\ell 0} & 0 \\
0 & 0 & 0 & \sin^2\theta Y_{\ell 0} \\
\end{pmatrix}\,,
\end{equation}

\begin{equation}
f^{\ell 0}_{ab} = \frac{r^2}{\sqrt{2\ell(\ell +1)(\ell -1)(\ell +2)}} \begin{pmatrix}
0 & 0 & 0 & 0 \\
0 & 0  & 0 & 0 \\
0 & 0 & W_{\ell 0} & 0 \\
0 & 0 & 0 & -\sin^2\theta W_{\ell 0} \\ 
\end{pmatrix}\,,
\end{equation}
where $Y^{\ell 0}$ are the $m=0$ spherical harmonics and $W^{\ell 0}$ are given by
\begin{equation}
W^{\ell 0}\equiv\left(\frac{d^2}{d\theta^2}-\cot\theta\frac{d}{d\theta}\right)Y^{\ell 0}.
\end{equation}

The coefficients of the source after a tensor spherical harmonics decomposition are
\allowdisplaybreaks[4]
\begin{widetext}
\begin{align}
A_{00}^{(0)}(r)=&-24\sqrt{\pi}\zeta\frac{m^4}{r^6}\frac{\chi^2}{f^2}\left(1-\frac{101}{18}\frac{m}{r}+\frac{25m^2}{r^2}-\frac{877}{18}\frac{m^3}{r^3}-\frac{1022}{15}\frac{m^4}{r^4}-\frac{2224}{9}\frac{m^5}{r^5}+\frac{107786}{45}\frac{m^6}{r^6}-\frac{53452}{15}\frac{m^7}{r^7}\right.
\nonumber \\
&\left.-\frac{208}{45}\frac{m^8}{r^8}+\frac{5920}{3}\frac{m^9}{r^9}\right),
\\
A_{20}^{(0)}(r)=&-\frac{44\sqrt{5\pi}}{15}\zeta\frac{m^5}{r^7}\chi^2\left(1+\frac{7737}{110}\frac{m}{r}-\frac{4201}{55}\frac{m^2}{r^2}+\frac{1047}{11}\frac{m^3}{r^3}-\frac{22086}{11}\frac{m^4}{r^4}+\frac{194424}{55}\frac{m^5}{r^5}-\frac{8880}{11}\frac{m^6}{r^6}\right),
\\
A_{00}(r)=&-2\sqrt{\pi}\zeta\frac{m^2}{r^4}\frac{\chi^2}{f^4}\left(1-\frac{4m}{r}+\frac{16}{3}\frac{m^2}{r^2}+\frac{40}{3}\frac{m^3}{r^3}-\frac{236}{3}\frac{m^4}{r^4}+\frac{482}{3}\frac{m^5}{r^5}+\frac{13672}{15}\frac{m^6}{r^6}-\frac{6416}{3}\frac{m^7}{r^7}-\frac{15288}{5}\frac{m^8}{r^8}\right.
\nonumber \\
&\left.+\frac{13808}{5}\frac{m^9}{r^9}+\frac{84928}{5}\frac{m^{10}}{r^{10}}-\frac{18560m^{11}}{r^{11}}\right),
\\
A_{20}(r)=&-\frac{236\sqrt{5\pi}}{75}\zeta\frac{m^4}{r^6}\frac{\chi^2}{f^2}\left(1+\frac{163}{59}\frac{m}{r}-\frac{2913}{118}\frac{m^2}{r^2}+\frac{3183}{59}\frac{m^3}{r^3}-\frac{4587}{59}\frac{m^4}{r^4}+\frac{25798}{59}\frac{m^5}{r^5}-\frac{59448}{59}\frac{m^6}{r^6}+\frac{34800}{59}\frac{m^7}{r^7}\right)
\\
B_{20}(r)=&\frac{92\sqrt{15\pi}}{75}\zeta\frac{m^4}{r^6}\frac{\chi^2}{f}\left(1+\frac{153}{46}\frac{m}{r}-\frac{651}{23}\frac{m^2}{r^2}-\frac{513}{23}\frac{m^3}{r^3}-\frac{1682}{23}\frac{m^4}{r^4}+\frac{33704}{69}\frac{m^5}{r^5}-\frac{2000}{23}\frac{m^6}{r^6}\right)
\\
G_{00}^{(s)}(r)=&\frac{56\sqrt{2\pi}}{3}\zeta\frac{m^4}{r^6}\frac{\chi^2}{f}\left(1-\frac{5}{4}\frac{m}{r}+\frac{333}{28}\frac{m^2}{r^2}+\frac{157}{14}\frac{m^3}{r^3}-\frac{969}{70}\frac{m^4}{r^4}-\frac{1807}{5}\frac{m^5}{r^5}+\frac{2068}{5}\frac{m^6}{r^6}+\frac{1320}{7}\frac{m^7}{r^7}\right),
\\
G_{20}^{(s)}(r)=&\frac{2\sqrt{10\pi}}{15}\zeta\frac{m^4}{r^6}\chi^2\left(1-\frac{15m}{r}-\frac{4963}{5}\frac{m^2}{r^2}-\frac{1164m^3}{r^3}-\frac{2910m^4}{r^4}+\frac{96528}{5}\frac{m^5}{r^5}-\frac{2640m^6}{r^6}\right),
\\
F_{20}(r)=&-\frac{4\sqrt{15\pi}}{15}\zeta\frac{m^4}{r^6}\chi^2\left(1+\frac{8m}{r}+\frac{809}{5}\frac{m^2}{r^2}-\frac{358}{5}\frac{m^3}{r^3}+\frac{1386}{5}\frac{m^4}{r^4}-\frac{25888}{15}\frac{m^5}{r^5}+\frac{2160m^6}{r^6}\right).
\end{align}
\\
Substituting these source terms into Eqs.~\eqref{dfq1}-\eqref{dfq5}, we obtain a set of ordinary differential equations for $H_{000}$, $H_{200}$, $K_{00}$, $H_{020}$, $H_{220}$, and $K_{20}$, which we solve to find
\\
\begin{align}
H_0^{00}(r)=&\frac{\sqrt{\pi}}{3}\zeta\frac{m^3}{r^3}\frac{\chi^2}{f}\left(1+\frac{14m}{r}+\frac{52}{5}\frac{m^2}{r^2}+\frac{1358}{15}\frac{m^3}{r^3}+\frac{652}{3}\frac{m^4}{r^4}+\frac{1204}{5}\frac{m^5}{r^5}-\frac{1792}{3}\frac{m^6}{r^6}-\frac{1120}{3}\frac{m^7}{r^7}\right),
\\
H_2^{00}(r)=&\sqrt{\pi}\zeta\frac{m^2}{r^2}\frac{\chi^2}{f^2}\left(1-\frac{m}{r}+\frac{10m^2}{r^2}-\frac{12m^3}{r^3}+\frac{218}{3}\frac{m^4}{r^4}+\frac{128}{3}\frac{m^5}{r^5}-\frac{724}{15}\frac{m^6}{r^6}-\frac{22664}{15}\frac{m^7}{r^7}+\frac{25312}{15}\frac{m^8}{r^8}+\frac{1600}{3}\frac{m^9}{r^9}\right),
\\
K_{00}(r)=&0,
\\
H_0^{20}(r)=&-\frac{17852\sqrt{5\pi}}{13125}\zeta\frac{m^3}{r^3}\frac{\chi^2}{f}\left(1+\frac{m}{r}+\frac{27479}{31241}\frac{m^2}{r^2}-\frac{2186945}{187446}\frac{m^3}{r^3}-\frac{448285}{31241}\frac{m^4}{r^4}-\frac{78975}{4463}\frac{m^5}{r^5}+\frac{1229650}{13389}\frac{m^6}{r^6}\right.
\nonumber \\
&\left.+\frac{303800}{13389}\frac{m^7}{r^7}-\frac{210000}{4463}\frac{m^8}{r^8}\right),
\\
H_2^{20}(r)=&-\frac{17852\sqrt{5\pi}}{13125}\zeta\frac{m^3}{r^3}\frac{\chi^2}{f}\left(1+\frac{3588}{4463}\frac{m}{r}-\frac{9271}{31241}\frac{m^2}{r^2}-\frac{7545095}{187446}\frac{m^3}{r^3}+\frac{1972315}{31241}\frac{m^4}{r^4}-\frac{446825}{4463}\frac{m^5}{r^5}+\frac{7215350}{13389}\frac{m^6}{r^6}\right.
\nonumber \\
&\left.-\frac{4809000}{4463}\frac{m^7}{r^7}+\frac{3570000}{4463}\frac{m^8}{r^8}\right),
\\
K_{20}(r)=&-\frac{8926\sqrt{10\pi}}{13125}\zeta\frac{m^3}{r^3}\chi^2\left(1+\frac{10370}{4463}\frac{m}{r}+\frac{266911}{62482}\frac{m^2}{r^2}+\frac{63365}{13389}\frac{m^3}{r^3}-\frac{309275}{31241}\frac{m^4}{r^4}-\frac{81350}{4463}\frac{m^5}{r^5}-\frac{443800}{13389}\frac{m^6}{r^6}
\right. \\
&\left. +\frac{210000}{4463}\frac{m^7}{r^7}\right).
\end{align}
\\
These solutions are then used to reconstruct the metric perturbation, as presented in the main text.

\section{High-Order Scalar Field}
\label{app2:high-o-field}
In this section, we present the scalar field to ${\cal{O}}(\chi^{8})$. Let us decompose the field as in Eq.~\eqref{vartheta-exp}, where $\vartheta^{(0,1)}$ 
was already presented in Eq.~\eqref{vartheta01} and $\vartheta^{(2,1)}$ was given in Eq.~\eqref{vartheta21}. Let us further define $\tilde{r} = r/M$
and $\tilde{\vartheta}^{(m,n)} = \vartheta^{(m,n)}/(\alpha/\beta)$. The nonvanishing, higher order pieces are then
\begin{align}
\tilde{\vartheta}^{(4,1)} &=-\frac{2}{35 \bar r^5}-\frac{1}{7 \bar r^4}-\frac{3}{14 \bar r^3}-\frac{1}{4
   \bar r^2}-\frac{1}{4 \bar r}+\left(\frac{360}{7 \bar r^7}+\frac{110}{7 \bar r^6}+\frac{22}{7 \bar r^5}\right)
   \cos ^4(\theta )+\left(\frac{4}{7 \bar r^6}+\frac{24}{35 \bar r^5}+\frac{3}{7
   \bar r^4}+\frac{1}{7 \bar r^3}\right) \cos ^2(\theta )\,,
\\
\tilde{\vartheta}^{(6,1)} &=-\frac{5}{252 \bar r^6}-\frac{5}{84 \bar r^5}-\frac{5}{48 \bar r^4}-\frac{5}{36
   \bar r^3}-\frac{5}{32 \bar r^2} -\frac{5}{32 \bar r}+\left(-\frac{896}{9 \bar r^9}-\frac{70}{3
   \bar r^8}-\frac{10}{3 \bar r^7}\right) \cos ^6(\theta )
\nn \\ 
&+  
   \left(-\frac{5}{6 \bar r^8}-\frac{5}{7 \bar r^7}-\frac{25}{84 \bar r^6}-\frac{5}{84 \bar r^5}\right) \cos
   ^4(\theta )+\left(\frac{1}{7 \bar r^7}+\frac{5}{21 \bar r^6}+\frac{3}{14
   \bar r^5}+\frac{1}{8 \bar r^4}+\frac{1}{24 \bar r^3}\right) \cos ^2(\theta
   )\,,
\\
\tilde{\vartheta}^{(8,1)} &=-\frac{1}{132 \bar r^7}-\frac{7}{264 \bar r^6}-\frac{7}{132 \bar r^5}-\frac{7}{88
   \bar r^4}-\frac{35}{352 \bar r^3}-\frac{7}{64 \bar r^2}-\frac{7}{64 \bar r}+\left(\frac{1800}{11
   \bar r^{11}}+\frac{342}{11 \bar r^{10}}+\frac{38}{11 \bar r^9}\right) \cos ^8(\theta
   )
\nn \\
&+\left(\frac{56}{55 \bar r^{10}}+\frac{112}{165 \bar r^9}+\frac{7}{33
   \bar r^8}+\frac{1}{33 \bar r^7}\right) \cos ^6(\theta )+\left(-\frac{2}{11
   \bar r^9}-\frac{5}{22 \bar r^8}-\frac{45}{308 \bar r^7}-\frac{5}{88 \bar r^6}-\frac{1}{88
   \bar r^5}\right) \cos ^4(\theta )
\nn \\
&+\left(\frac{1}{22 \bar r^8}+\frac{15}{154
   \bar r^7}+\frac{5}{44 \bar r^6}+\frac{1}{11 \bar r^5}+\frac{9}{176 \bar r^4}+\frac{3}{176
   \bar r^3}\right) \cos ^2(\theta) \,.
\end{align}
\ew
In deriving these expressions, we have required that the scalar field be asymptotically flat (at spatial infinity)
and regular at the Kerr event horizon. 

\bibliographystyle{apsrev}
\bibliography{biblio}
\end{document}